\documentclass[twocolumn,aps,showpacs]{revtex4}
\usepackage{epsfig}

\begin{document}
\voffset 1.8cm

\title{\LARGE \bf Microscopic description of the scissors \\ mode
   in odd-mass heavy deformed nuclei}

\author{Carlos E. Vargas}
\email{vargas@ganil.fr}
\affiliation{Grand Acc\'el\'erateur National d'Ions Lourds,
BP 5027, F-14076 Caen Cedex 5, France}
\author{Jorge G. Hirsch}
\email{hirsch@nuclecu.unam.mx}
\affiliation{Instituto de Ciencias Nucleares, Universidad Nacional
Aut\'onoma de M\'exico,Apartado Postal 70-543 M\'exico 04510 DF, M\'exico }
\author{Jerry P. Draayer}
\email{draayer@lsu.edu}
\affiliation{Department of Physics and Astronomy,
           Louisiana State University,
Baton Rouge, Louisiana 70803, U.S.A.}

\vskip 0.5cm
\date{\today}



\begin{abstract}

Pseudo-SU(3) shell-model results are reported for M1 excitation strengths in
$^{157}$Gd, $^{163}$Dy and $^{169}$Tm in the energy range between 2 and 4 MeV.
Non-zero pseudo-spin couplings between the configurations play a very 
important role in determining the M1 strength distribution, especially its 
rapidly changing fragmentation pattern which differs significantly from
what has been found in neighboring even-even systems. The results
suggest one should examine contributions from intruder levels.

\end{abstract}
\maketitle

The scissors mode in nuclei refers to a pictorial image of deformed
proton and neutron distributions oscillating against one another
\cite{Iud78}. A description of this mode within the framework of
the IBM \cite{Die83} led to its detection in $^{156}$Gd using
high-resolution inelastic electron scattering \cite{Boh84}.
Systematic studies employing nuclear resonance fluorescence scattering
(NRF) measurements \cite{Ber84} followed.
The non-observation of these low-energy M1 excitations in
inelastic proton scattering (IPS) \cite{Dja85} confirmed its
orbital character \cite{Wes89}. Over the past two decades an impressive
wealth of information about the scissors mode in even-even nuclei
has been obtained and analyzed \cite{Kne96}.

Low-energy M1 transitions in odd-mass nuclei were first observed
in $^{163}$Dy in 1993 \cite{Bau93}. Unambiguous spin and parity
assignments of excited states in these nuclei are difficult
to make due to the half-integer character of the angular momentum of the
states \cite{Mar95}. Furthermore, the M1 strengths in odd-mass nuclei 
are highly
fragmented. Since the intensities are far smaller than in even-even
nuclei, their identification against the background \cite{End97}, which is
complicated by the presence of a small amount of impurities in the target
\cite{Kne96}, requires much higher experimental resolution \cite{Fri94}.

Theoretical descriptions of scissors modes in odd-mass nuclei
have been offered within the context of the IBFM \cite{Van89,Fra91},
the particle-core-coupling model \cite{Rad89} and the QPNM \cite{Sol96}.
While the different models agree in relating the presence of the
uncoupled nucleon with the observed fragmentation,
the detailed description of this mode,
with a nearly flat spectrum in some nuclei and has
well-defined peaks in others is still not understood.
Recently, the interplay between the spin and orbital M1 channels was examined
\cite{Fay00} in the energy range between 4-10 MeV \cite{Iud01}.

In the present letter we analyze scissors-like M1 transitions in 
$^{157}$Gd, $^{163}$Dy
and $^{169}$Tm. These nuclei have been studied experimentally by a 
number of researchers
\cite{Bau93,Mar95,Hux99}. A fully microscopic description of M1 
transitions strengths
between 2 and 4 MeV in these rare-earth nuclei was carried out using 
the pseudo SU(3)
shell model. Good qualitative descriptions of the fragmentation of 
the M1 transition
strength is obtained by including, for the first time, states with 
pseudo-spin 1 (in
addition to $\tilde{S}$ = 0) and 3/2 (in addition to $\tilde{S}$ = 
1/2). For normal
parity levels our findings suggest that while orbital couplings are 
important, in
odd-even mass nuclei it is spin-flip type couplings that dominate M1 
strengths in the
low-energy domain.  These spin-flip type transitions were also found 
to be essential
for describing the rapidly changing fragmentation patterns found in 
neighboring odd-A
nuclei. Freezing the unique parity orbitals, which is the usual 
assumption, prevents
the theory from giving a quantitative description of the M1 strength, 
a result that
is not surprising since intruder states have the largest $l$ values 
and therefore
contribute maximally to orbital-type M1 transitions.

The pseudo SU(3) model \cite{Hec69,Ari69} capitalizes on the 
existence of pseudo-spin
symmetry, which refers to the experimental fact that single-particle 
orbitals with $j$
= $l$ - 1/2 and $j$ = ($l$ - 2) + 1/2 in the shell $\eta$ lie very 
close in energy and
can therefore be labeled as pseudo-spin doublets with quantum numbers 
$\tilde{j}$ =
$j$, $\tilde{\eta}$ = $\eta$ - 1, and $\tilde{l}$ = $l$ - 1. Its 
origin has been traced
back to the relativistic Dirac equation \cite{Blo95}. In the present 
version of the
pseudo-SU(3) model, the intruder level with opposite parity in each 
major shell is
removed from active consideration \cite{Var98} and pseudo-orbital and 
pseudo-spin
quantum numbers are assigned to the remaining single-particle states. 
This assumption
represents the strongest limitation of the present model.

Many-particle states of $n_\alpha$ active nucleons
($\alpha = p, n$) in a given ($N$) normal
parity shell $\eta_\alpha^N$ are classified by the following group chain
\cite{Dra82,Cas87,Var00}:

\begin{eqnarray}
~ \{ 1^{n^{N}_\alpha} \} ~~~~~~~ \{ \tilde{f}_\alpha \} ~~~\{ f_\alpha
\} ~\gamma_\alpha ~~~ (\lambda_\alpha , \mu_\alpha ) ~~~ \tilde{S}_\alpha
~~ K_\alpha  \nonumber \\
U(\Omega^N_\alpha ) \supset U(\Omega^N_\alpha / 2 ) \times U(2) \supset
SU(3) \times SU(2) \supset \nonumber \\
\tilde{L}_\alpha  ~~~~~~~~~~~~~~~~~~~~~ J_\alpha ~~~~ \nonumber \\
SO(3) \times SU(2) \supset SU_J(2),
\label{eq:chains}
\end{eqnarray}

\noindent where above each group the quantum numbers that characterize its
irreducible representations (irreps) are given and $\gamma_\alpha$ and
$K_\alpha$ are multiplicity labels of the indicated reductions.

The Hamiltonian used in the calculations includes spherical Nilsson
single-particle terms for the protons and neutrons ($H_{sp,\pi[\nu]}$), the
quadrupole-quadrupole ($\tilde Q \cdot \tilde Q$) and pairing
($H_{pair,\pi[\nu]}$) interactions, as well as three `rotor-like' terms
that are diagonal in the SU(3) basis:

\begin{eqnarray}
    H & = & H_{sp,\pi} + H_{sp,\nu} - \frac{1}{2}~ \chi~ \tilde Q \cdot
            \tilde Q - ~ G_\pi ~H_{pair,\pi} ~\label{eq:ham} \\
      &   & - ~G_\nu ~H_{pair,\nu} + ~a~ K_J^2~ +~ b~ J^2~ +~ A_{sym}~
            \tilde C_2 . \nonumber
\end{eqnarray}

\noindent A detailed analysis of each term of this Hamiltonian and its
parametrization can be found in \cite{Var00}. The three free parameters
$a, b, A_{sym}$ were fixed by the best reproduction of the low-energy
spectra; no additional parameters enter into the theory -- the calculated
M1 transitions reported below were not fit to the data.

A description of the low-energy spectra and B(E2) transition 
strengths in even-even
nuclei \cite{Pop00} and odd-mass heavy  deformed nuclei 
\cite{Var00,Var01} have been
carried out using linear combinations of SU(3) coupled proton-neutron 
irreps with
largest $C_2$ values and pseudo-spin 0 and 1/2 (for even and odd 
number of nucleons,
respectively), which are mixed by the single-particle terms in the Hamiltonian.

The large number of states that can decay through M1 transitions to 
the ground state in
odd-mass nuclei, led us to enlarge the basis by including states with 
pseudo-spin 1 and
3/2. These configurations are necessary to describe excited 
rotational bands and to
account for the strong fragmentation of the M1 strengths between 2 and 4 MeV in
odd-mass nuclei.

The inclusion of configurations with pseudo-spin 1 and 3/2 in the Hilbert
space allows for a description of highly excited rotational bands in
odd-mass nuclei.  This is illustrated in Ref. \cite{Var02}, where several
rotational bands in $^{157}$Gd, $^{163}$Dy and $^{169}$Tm are described,
including both excitation energies and intra- and inter-band B(E2)
transition strengths, and shown to be in close agreement with the
experimental data. In contrast, when the configuration space was
restricted to the most spatially symmetric configurations, those with
pseudo-spin 0 and 1/2, it was only possible to describe in $^{163}$Dy the
first three low-energy bands \cite{Var01}. The pseudo-spin symmetry is still
approximately preserved in the present case, with these three low-energy
bands showing only a small amount of pseudo-spin 1 and  3/2 admixing into
predominantly pseudo-spin 0 and 1/2 configurations, respectively.

The M1 transitions are mediated by the operator

\begin{equation}
T^1_\mu = \sqrt{\frac {3}{4\pi}} \mu_N \{ g^o_\pi L^\pi_\mu +
g^S_\pi S^\pi_\mu + g^o_\nu L^\nu_\mu + g^S_\nu S^\nu_\mu \}
\label{eq:t1op}
\end{equation}
\noindent with
\begin{equation}
L^{\pi,[\nu]} = \sum_i^{Z,[N]} l^{\pi,[\nu]} (i) ~,~~
S^{\pi,[\nu]} = \sum_i^{Z,[N]} s^{\pi,[\nu]} (i) .
\end{equation}
\noindent In Eq.  (\ref{eq:t1op}) the
orbital and `quenched' (by a factor of 0.7) spin $g$ factors for 
protons and neutrons
are used:
\begin{equation}
g^o_\pi = 1,~~g^o_\nu = 0,~~g^S_\pi = (0.7) 5.5857,~~g^S_\nu = - (0.7) 3.8263 .
\label{eq:m1para}
\end{equation}

\noindent To evaluate the M1 transition operator between eigenstates of
the Hamiltonian (\ref{eq:ham}), the pseudo SU(3) tensorial expansion of
the T1 operator (\ref{eq:t1op}) \cite{Cas87} was employed.

In what follows, the B(M1;$J_i^\pi \rightarrow J_f^\pi$) transitions in
$^{157}$Gd, $^{163}$Dy and $^{169}$Tm are presented. $J_i^\pi$ refers to
the ground states $3/2^-$, $5/2^-$ and $1/2^+$ in these nuclei.
In each figure, insert a) corresponds to the experimental
results, while insert b) represents the theoretical values
obtained with the T1 operator of Eq. (\ref{eq:t1op}). Insert c) shows the
values with $g^o_{\pi,\nu}$ in Eq. (\ref{eq:t1op}) set to zero, i.e.
with only the spin part of the T1 operator taken into account, and insert
d) shows the results with $g^s_{\pi,\nu}$ in Eq. (\ref{eq:t1op}) set to
zero, i.e. including only the orbital part of T1.

\begin{figure}
\vspace*{-1.9cm}
\hspace{-0.7cm}
\epsfxsize=9.34cm
\centerline{\epsfbox{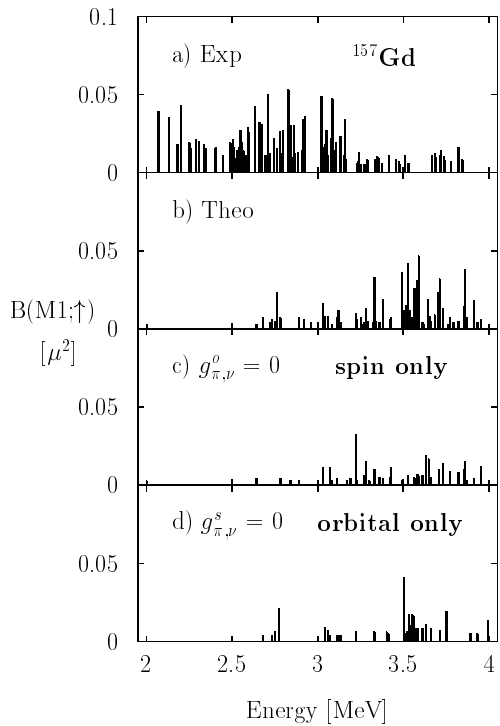}}\vspace{-1cm}
\caption{Distribution of M1 transitions between 2 and 4 MeV for
$^{157}$Gd. Insert a)
shows the experimental values \cite{Mar95}, insert b) shows the theoretical
with the complete T1 operator, insert c) shows the values with
$g^o_{\pi,\nu}$ = 0 (only the spin channel) and insert d) with
$g^s_{\pi,\nu}$ = 0 (only the orbital channel).}
\label{m1-gd}
\end{figure}

\begin{figure}
\vspace*{-1.9cm}
\hspace{-0.7cm}
\epsfxsize=9.34cm
\centerline{\epsfbox{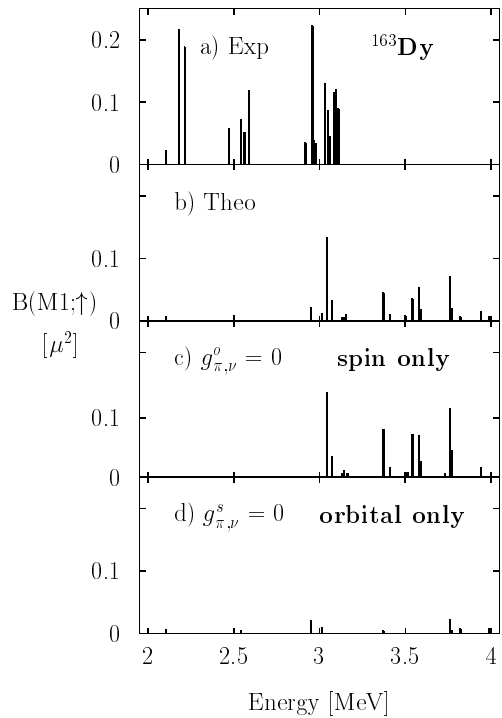}}
\vspace{-1cm}
\caption{M1 transitions for $^{163}$Dy. Convention is the same as in Figure
\ref{m1-gd}. Experimental values taken from Ref. \cite{Bau93}.}
\label{m1-dy}\end{figure}

\begin{figure}
\vspace*{-1.9cm}
\hspace{-0.7cm}
\epsfxsize=9.34cm
\centerline{\epsfbox{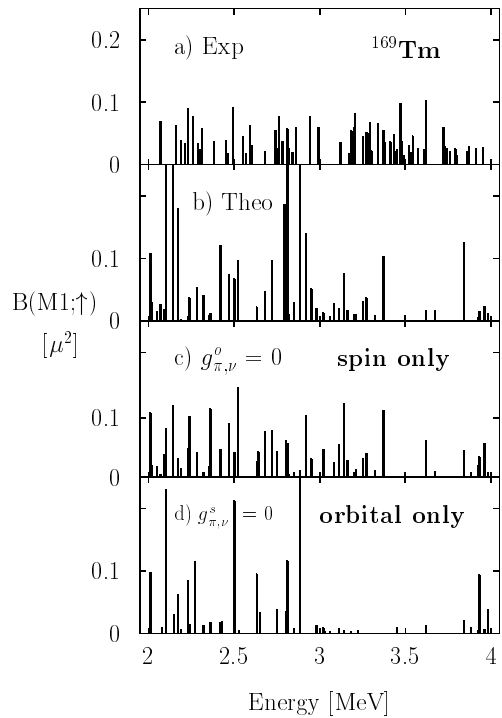}}\vspace{-1cm}
\caption{M1 transitions for $^{169}$Tm. Convention is the same as in Figure
\ref{m1-gd}. Experimental values were taken from Ref. \cite{Hux99}.}
\label{m1-tm}
\end{figure}

The differences between the M1 transition strength distribution in
$^{157}$Gd, $^{163}$Dy and $^{169}$Tm, shown Figs.
\ref{m1-gd}, \ref{m1-dy}, and \ref{m1-tm} respectively (notice the change on
the scale), are both striking and well-known \cite{End97}.
In $^{157}$Gd there are 88 known M1 transitions between 2 and 4 MeV,
all smaller than 0.05 $\mu^2$ and distributed in a nearly
flat spectrum. In $^{163}$Dy the M1 transition strengths are
distributed only among 17 peaks, clustered in three well-defined
groups, and most of them have strengths between 0.1 and 0.2 $\mu^2$.
$^{169}$Tm has an intermediate degree of fragmentation, with some
clustered structures and many transitions on the order of
0.1 $\mu^2$.

\begin{table}
\begin{tabular}{ccccc}
   $^{157}$Gd & E $<$ 2 MeV & 2 - 4 MeV & 4 MeV $<$ E \\
Experiment \cite{Mar95}  & & 1.596  $\pm$0.235  	&    \\
Theory       &  0.232    & 0.782     		& 0.613 \\
Spin only    &  0.138    & 0.389 	      	& 0.371 \\
Orbital only &  0.084	 & 0.308	      	& 0.385 \\ ~~\\
$^{163}$Dy & E $<$ 2 MeV & 2 - 4 MeV & 4 MeV $<$ E \\
Experiment \cite{Bau93}  & & 1.641 $\pm$ 0.338  	&    \\
Theory       &  0.630    & 0.483              	& 0.030 \\
Spin only    &  0.908    & 0.543 	      	& 0.026 \\
Orbital only &  0.088	 & 0.103	      	& 0.012  \\ ~~\\
$^{169}$Tm & E $<$ 2 MeV & 2 - 4 MeV & 4 MeV $<$ E \\
Experiment \cite{Hux99}  & 1.912 $\pm$0.244 & 2.833 $\pm$0.812  &
0.515 $\pm$0.274  \\
Theory       &  2.460    & 3.769               	& 0.435 \\
Spin only    &  2.245    & 2.332 	      	& 0.164 \\
Orbital only &  1.483	 & 1.838	      	& 0.321
\end{tabular}
\caption{Summed B(M1;$\uparrow$) strengths (in $\mu^2$) in the
different energy regions.}
\end{table}

Using an enlarged version of pseudo SU(3) shell-model theory described
above, we obtained a microscopic description of these M1 transitions
and their fragmentation in the three nuclei.
The gross features of the M1 strength distributions in each
of the nuclei are clearly reproduced, i.e. the different 
fragmentation patterns.
On the other hand, for $^{157}$Gd and $^{163}$Dy the M1 strength
distributions are displaced toward higher energies by about 0.75 MeV and
the total sums are underestimated.
This effect
could be related with the absence of spin dependent terms in the Hamiltonian
(\ref{eq:ham}). For $^{169}$Tm the distribution in energy of the M1 strengths
is correct, but some transition strengths are overestimated by a factor
2 to 3.

The ground state wave functions of the two nuclei with odd number of
neutrons, $^{157}$Gd and $^{163}$Dy, have one important difference. In
$^{163}$Dy the ground state has only pseudo-spin 0 and 1/2 components,
while $^{157}$Gd has a 13\% mixing with pseudo-spin 1 and 3/2 
components. In the M1
transition matrix elements the presence of these components in the 
later case gives
rise to interference and fragmentation, while its absence in the 
former nuclei is
associated with few large M1 transitions. The odd proton number of 
$^{169}$Tm allows
orbital proton excitations between half-integer components, building 
up its large M1
summed transition strength.

Having analyzed the similarities and differences between the
experimental data and the theoretical predictions, we proceed to discuss
the spin and orbital contributions to the M1 transitions.
In insert c) of each figure the M1 transition strengths calculated
only with the spin operators, i.e. making $g^o_{\pi,\nu} =
0$ in Eq. \ref{eq:t1op}), is presented.
Insert d)  shows the M1 strength when only the orbital
part of the operator (\ref{eq:t1op}) are included ($g^s_{\pi,\nu} = 0$).
In all cases the spin coupling is by far the dominant mode,
but for $^{169}$Tm the orbital contribution is also large.

In the case of $^{163}$Dy, there is an almost null contribution from the
orbital part of the transition operator (0.103 $\mu^2$), which in fact
interferes destructively with the spin channel (0.543 $\mu^2$) to produce
a summed M1 strength of 0.483 $\mu^2$ in the scissors energy region.

The `angle' between the orbital and spin channels, as defined by
Fayache {\it et al.} \cite{Fay00} is 110$^o$ for $^{163}$Dy. For $^{157}$Gd,
this angle has a value of 83$^o$ and for $^{169}$Tm it is 96$^o$.
 From Table I it can be seen that below 2 MeV
the spin transitions are clearly dominant.
Nevertheless, it should be emphasized that contributions of 
the intruder sector have been neglected.

The pseudo SU(3) shell model for odd-mass nuclei has been shown to 
offer a qualitative
microscopic description of the scissors mode and its fragmentation. In order to
successfully reproduce the observed fragmentation of the M1 strength, 
it was necessary
to use realistic values for the single particle energies and to 
enlarge the Hilbert
space to include those pseudo SU(3) irreps with the largest $C_2$ values and
pseudo-spin 1 and 3/2. This expansion of the model space allowed the 
T1 operator to
connect the ground state with many excited states ($|J_f - J_i| \leq 1$) in the
energy range between 2 and 4 MeV. The transitions are dominated by 
spin couplings, but
interference with the orbital mode is very important.

A fully quantitative treatment of the problem should take into 
account contribution
from the intruder sector.  Detailed studies of M1 transitions in other odd-mass
nuclei are under investigation and should offer an opportunity to 
further apply and
test the theory.

The authors thank S. Pittel, A. Frank and P. Van Isacker for 
constructive comments.
This work was supported in part by CONACyT (M\'exico) and the US 
National Science
Foundation.



\end{document}